\documentclass[aps,twocolumn,prb,showpacs,preprintnumbers,amsmath,amssymb,superscriptaddress]{revtex4}

\usepackage{graphicx,epsfig}
\usepackage{dcolumn}
\usepackage{bm}

\begin{document}

\title{Fully Gapped Single-Particle Excitations in the Lightly Doped
 Cuprates}

\author{K.M. Shen}
 \affiliation{Departments of Applied Physics, Physics, and Stanford
 Synchrotron Radiation Laboratory, Stanford University, Stanford,
 California 94305}

\author{T. Yoshida}
 \affiliation{Departments of Applied Physics, Physics, and Stanford
 Synchrotron Radiation Laboratory, Stanford University, Stanford,
 California 94305}

\author{D.H. Lu}
 \affiliation{Departments of Applied Physics, Physics, and Stanford
 Synchrotron Radiation Laboratory, Stanford University, Stanford,
 California 94305}

\author{F. Ronning}
 \altaffiliation[Present address: ]{Los Alamos National Laboratory, Los
 Alamos, New Mexico}
 \affiliation{Departments of Applied Physics, Physics, and Stanford
 Synchrotron Radiation Laboratory, Stanford University, Stanford,
 California 94305}

\author{N.P. Armitage}
 \altaffiliation[Present address: ]{Department of Physics and Astronomy,
 University of California, Los Angeles, Los Angeles, CA}
 \affiliation{Departments of Applied Physics, Physics, and Stanford
 Synchrotron Radiation Laboratory, Stanford University, Stanford,
 California 94305}

\author{W.S. Lee}
 \affiliation{Departments of Applied Physics, Physics, and Stanford
 Synchrotron Radiation Laboratory, Stanford University, Stanford,
 California 94305}

\author{X.J. Zhou}
 \affiliation{Departments of Applied Physics, Physics, and Stanford
 Synchrotron Radiation Laboratory, Stanford University, Stanford,
 California 94305}

\author{A. Damascelli}
 \altaffiliation[Present address: ]{Department of Physics and Astronomy, University of British Columbia,
 Vancouver, B.C., Canada}
 \affiliation{Departments of Applied Physics, Physics, and Stanford
 Synchrotron Radiation Laboratory, Stanford University, Stanford,
 California 94305}

\author{D.L. Feng}
 \altaffiliation[Present address: ]{Department of Physics and Synchrotron Radiation Research
 Center, Fudan University, Shanghai, China}
 \affiliation{Departments of Applied Physics, Physics, and Stanford
 Synchrotron Radiation Laboratory, Stanford University, Stanford,
 California 94305}

\author{N.J.C. Ingle}
 \affiliation{Departments of Applied Physics, Physics, and Stanford
 Synchrotron Radiation Laboratory, Stanford University, Stanford,
 California 94305}

\author{H. Eisaki}
 \affiliation{Departments of Applied Physics, Physics, and Stanford
 Synchrotron Radiation Laboratory, Stanford University, Stanford,
 California 94305}

\author{Y. Kohsaka}
 \affiliation{Departments of Physics, Complexity Science and
 Engineering, and Advanced Materials Science,
 University of Tokyo, Bunkyo-ku, Tokyo 113-0033, Japan}

\author{H. Takagi}
 \affiliation{Departments of Physics, Complexity Science and
 Engineering, and Advanced Materials Science,
 University of Tokyo, Bunkyo-ku, Tokyo 113-0033, Japan}

\author{T. Kakeshita}
 \affiliation{Departments of Physics, Complexity Science and
 Engineering, and Advanced Materials Science,
 University of Tokyo, Bunkyo-ku, Tokyo 113-0033, Japan}

\author{S. Uchida}
 \affiliation{Departments of Physics, Complexity Science and
 Engineering, and Advanced Materials Science,
 University of Tokyo, Bunkyo-ku, Tokyo 113-0033, Japan}

\author{P.K. Mang}
 \affiliation{Departments of Applied Physics, Physics, and Stanford
 Synchrotron Radiation Laboratory, Stanford University, Stanford,
 California 94305}

\author{M. Greven}
 \affiliation{Departments of Applied Physics, Physics, and Stanford
 Synchrotron Radiation Laboratory, Stanford University, Stanford,
 California 94305}

\author{Y. Onose}
 \affiliation{Departments of Physics, Complexity Science and
 Engineering, and Advanced Materials Science,
 University of Tokyo, Bunkyo-ku, Tokyo 113-0033, Japan}

\author{Y. Taguchi}
 \affiliation{Departments of Physics, Complexity Science and
 Engineering, and Advanced Materials Science,
 University of Tokyo, Bunkyo-ku, Tokyo 113-0033, Japan}

\author{Y. Tokura}
 \affiliation{Departments of Physics, Complexity Science and
 Engineering, and Advanced Materials Science,
 University of Tokyo, Bunkyo-ku, Tokyo 113-0033, Japan}

 \author{Seiki Komiya}
 \affiliation{Central Research Institute of Electric Power
 Industry, Komae, Tokyo 201-8511, Japan}

\author{Yoichi Ando}
 \affiliation{Central Research Institute of Electric Power
 Industry, Komae, Tokyo 201-8511, Japan}

\author{M. Azuma}
 \affiliation{Institute for Chemical Research, Kyoto University, Uji, Kyoto
611-0011, Japan}

\author{M. Takano}
 \affiliation{Institute for Chemical Research, Kyoto University, Uji, Kyoto
611-0011, Japan}

\author{A. Fujimori}
 \affiliation{Departments of Physics, Complexity Science and
 Engineering, and Advanced Materials Science,
 University of Tokyo, Bunkyo-ku, Tokyo 113-0033, Japan}

\author{Z.-X. Shen}
 \affiliation{Departments of Applied Physics, Physics, and Stanford
 Synchrotron Radiation Laboratory, Stanford University, Stanford,
 California 94305}

\date{\today}

\begin{abstract}

The low-energy excitations of the lightly doped cuprates were
studied by angle-resolved photoemission spectroscopy. A finite gap
was measured over the entire Brillouin zone, including along the
$d_{x^2 - y^2}$ nodal line. This effect was observed to be generic
to the normal states of numerous cuprates, including hole-doped
La$_{2-x}$Sr$_{x}$CuO$_{4}$ and
Ca$_{2-x}$Na$_{x}$CuO$_{2}$Cl$_{2}$ and electron-doped
Nd$_{2-x}$Ce$_{x}$CuO$_{4}$. In all compounds, the gap appears to
close with increasing carrier doping. We consider various
scenarios to explain our results, including the possible effects
of chemical disorder, electronic inhomogeneity, and a competing
phase.

\end{abstract}




\maketitle

The parent compounds of the cuprates are half-filled
antiferromagnetic insulators whose Coulomb repulsion opens a large
charge-transfer gap ($\sim$ 2 eV). While accurately describing the
single particle excitations of the undoped insulator remains a
theoretical challenge, this problem becomes far more daunting upon
the addition of even a small number of holes or electrons. The
exotic properties exhibited by these underdoped cuprates have led
to numerous inquiries and debates over the physics of the
insulator-superconductor transition, the presence of competing
phases, precursor superconductivity, and electronic phase
separation. While popular theoretical models (i.e. $t\!-\!J$ or
Hubbard models) predict the formation of metallic states even at
infinitesimally small doping concentrations \cite{Dagotto94},
antiferromagnetic N\'{e}el order has been experimentally found to
persist up to finite doping levels. Moreover, the doping range in
which low temperature insulating behavior is observed has been
universally found in the cuprates to extend well past the
disappearance of N\'{e}el order \cite{Iye92}. This naturally
raises the question of the respective roles played by order (i.e.
N\'{e}el order or alternative competing orders) and disorder
(chemical or electronic inhomogeneity), on the low lying
electronic states derived from doping the parent insulator.

To date, angle-resolved photoemission spectroscopy (ARPES) has
been the premier tool for the direct study of the electronic
structure of the near-optimally doped cuprates
\cite{Damascelli03}. However, its contributions to our
understanding of the lightly doped regime have been extremely
limited. The main reason for this disparity is that the bismuth
based cuprates, the archetypal materials for ARPES, are naturally
grown within a limited range around optimal doping, and thus it
becomes necessary to study other families in order to access
lighter dopings. Because of the extreme dearth of ARPES data in
the lightly doped regime, a serious gap exists in our experimental
understanding of the doping evolution of the electronic structure.
Unfortunately, this deficiency in knowledge occurs where the
physics of the cuprates is generally acknowledged to be most
complex, further complicating attempts at understanding the
physics of high-temperature superconductivity. It has been very
recently shown that upon the addition of even a small
concentration of carriers to the parent insulator, finite spectral
weight develops near the chemical potential
\cite{Armitage02,Yoshida02}, as expected for a compressible
thermodynamic system. While consistent with the high temperature
metallic behavior seen at low concentrations \cite{Iye92,Ando01},
this near-E$_{\mathrm{F}}$ weight has serious conflicts with the
low temperature insulating behavior observed by charge and thermal
transport \cite{Iye92,Ando01,Hawthorn03,Sun03}. It is therefore
important to ask whether this near-E$_{\mathrm{F}}$ weight is
additionally gapped at low energies. An energy gap along the
$d$-wave node has not yet been observed, as previous work in the
pseudogap regime was restricted to higher dopings where the nodal
states were found to be ungapped. The presence of a gap along the
nodal direction will clearly demonstrate the effects of disorder
or additional orders on the ostensibly $d$-wave-like low energy
states.

Here we report an extensive ARPES study of various lightly doped
cuprates, where we find an apparently finite excitation gap in the
normal state over the entire Brillouin zone. This result is
observed for a variety of compounds and carrier concentrations,
including hole-doped Ca$_{2-x}$Na$_{x}$CuO$_{2}$Cl$_{2}$ (\emph{x}
= 0.05, 0.10), La$_{2-x}$Sr$_{x}$CuO$_{4}$ (\emph{x} = 0.01,
0.02), and electron-doped Nd$_{2-x}$Ce$_{x}$CuO$_{4}$ (\emph{x} =
0.025, 0.04). This study reports the lowest doping concentrations
ever studied by ARPES for each of the respective compounds,
including La$_{2-x}$Sr$_{x}$CuO$_{4}$ with $x \leq 0.02$, where
N\'{e}el order persists. The widespread absence of ungapped
excitations in this regime suggests that this behavior may be
generic to the cuprate superconductors instead of being a
material-specific phenomenon. We consider a number of scenarios,
including the presence of disorder and electronic inhomogeneity,
as well as a possible competing order.

ARPES measurements were performed at Beamline 5-4 of the Stanford
Synchrotron Radiation Laboratory with both synchrotron radiation
and a He discharge lamp in conjunction with a Scienta SES-200
electron analyzer operating in parallel angle detection mode. The
typical energy and angular resolutions used for these measurements
were between 10 to 14 meV and 0.3$^{\circ}$, respectively. Even
with an energy resolution of 14 meV, edge positions could be
measured accurately and reproducibly to within 1 meV. The Fermi
energy (E$_{\mathrm{F}}$) was determined from a polycrystalline Au
target in direct electrical contact with the sample. Single
crystals of La$_{2-x}$Sr$_{x}$CuO$_{4}$ and
Nd$_{2-x}$Ce$_{x}$CuO$_{4}$ were grown using the
travelling-solvent floating zone method, while single crystals of
Ca$_{2-x}$Na$_{x}$CuO$_{2}$Cl$_{2}$ were grown using a self-flux
method \cite{KohsakaGrowth02}. Samples were first aligned by Laue
diffraction \emph{ex situ}, and cleaved and measured at a base
temperature of 15 K at a pressure of better than 5$\times10^{-11}$
torr.

\begin{figure}[t]
\includegraphics{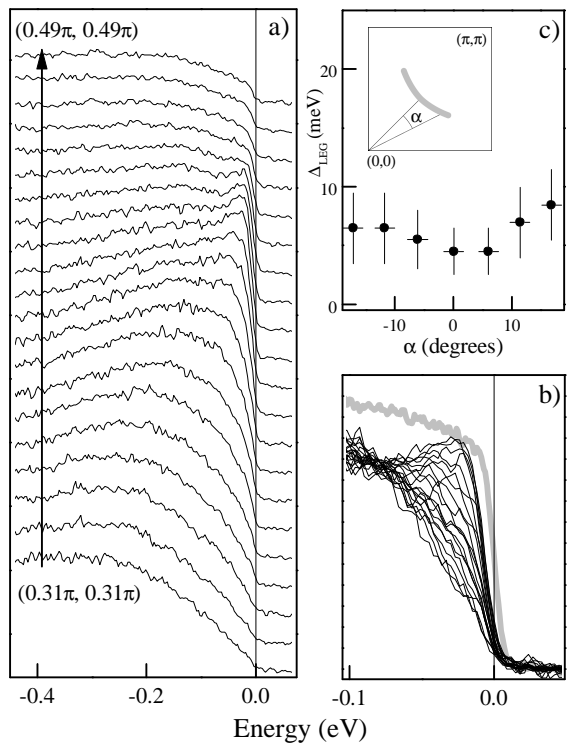} \vspace{0in}
\caption{(a) Spectra from Ca$_{1.9}$Na$_{0.1}$CuO$_{2}$Cl$_{2}$
taken along (0,0) to ($\pi$,$\pi$). (b) Collapsed spectra from (a)
with a finite leading edge gap of 5 meV as deduced from comparison
with Au (grey). (c) Momentum dependence of this leading edge gap
along the low-energy arc as a function of angle $\alpha$.}
\label{EDCCut}
\end{figure}

As previously observed in La$_{2-x}$Sr$_{x}$CuO$_{4}$ and
Bi$_2$Sr$_2$CaCu$_2$O$_{8+\delta}$ \cite{Marshall96,Yoshida02},
the first hole addition states emerge near
($\frac{\pi}{2}$,$\frac{\pi}{2}$), the top of the lower Hubbard
band of the undoped parent insulator \cite{Wells95} (for a
comprehensive overview, see Ref. \onlinecite{Damascelli03}). At
relatively low concentrations, the locus of low-lying spectral
weight is confined to a discontinuous arc. Outside this sector, a
large pseudogapped region devoid of well-defined low energy
excitations persists around ($\pi$,0). As an example, we show
spectra from Ca$_{1.9}$Na$_{0.1}$CuO$_{2}$Cl$_{2}$ in Figure
\ref{EDCCut} whose electronic structure has been shown to be
consistent with the behavior described above
\cite{Ronning03,KohsakaARPES02}. A dispersive excitation branch
can be observed along the (0,0)-($\pi$,$\pi$) line in Figure
\ref{EDCCut}a. In Figure \ref{EDCCut}b, all spectra have been
collapsed together and there exists a clear shift of all leading
edge midpoints away from the chemical potential which we call a
leading edge gap (LEG). As mentioned above, the locus of low-lying
excitations in this compound is confined to an arc-like segment
spanning approximately $\pm$20$^{\circ}$ measured radially from
(0,0). The angular dependence of this gap,
$\Delta_{\mathrm{LEG}}$, shown in Figure \ref{EDCCut}c, exhibits
weak anisotropy within this arc. However, it is difficult to
ascertain whether this apparent anisotropy is intrinsic, or due to
the loss of spectral weight and broader lineshapes away from the
nodal line, as illustrated by the larger error bars.

\begin{figure}[t]
\includegraphics{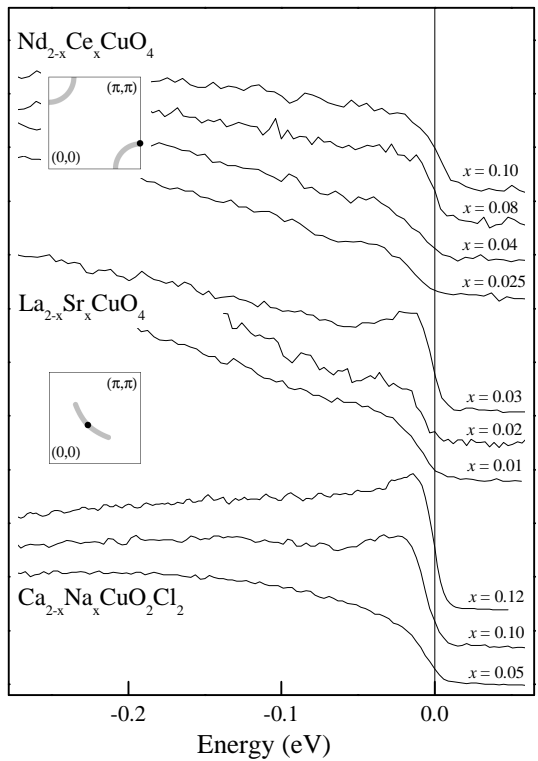} \vspace{0in}
\caption{LEG spectra from hole-doped
Ca$_{2-x}$Na$_{x}$CuO$_{2}$Cl$_{2}$ and
La$_{2-x}$Sr$_{x}$CuO$_{4}$, and electron-doped
Nd$_{2-x}$Ce$_{x}$CuO$_{4}$. The bottom inset shows the wavevector
of the spectra taken from Ca$_{2-x}$Na$_{x}$CuO$_{2}$Cl$_{2}$ and
La$_{2-x}$Sr$_{x}$CuO$_{4}$; the top shows
Nd$_{2-x}$Ce$_{x}$CuO$_{4}$. All data were taken at 15 K.}
\label{EDCSummary}
\end{figure}

Spectra from non-superconducting (\emph{x} = 0.05) and underdoped
(\emph{x} = 0.10, T$_{\mathrm{c}}$ = 13 K; \emph{x} = 0.12,
T$_{\mathrm{c}}$ = 22 K) compositions of
Ca$_{2-x}$Na$_{x}$CuO$_{2}$Cl$_{2}$ are shown at the bottom of
Figure \ref{EDCSummary}. While no well-defined peak is visible for
\emph{x} = 0.05, there exists a distinct edge structure with
$\Delta_{\mathrm{LEG}}$ = 7 meV. This effect decreases with hole
doping and appears to close by \emph{x} = 0.12. To demonstrate
that this effect is generic to all cuprates, we also present
results from very lightly doped, non-superconducting
La$_{2-x}$Sr$_{x}$CuO$_{4}$ and Nd$_{2-x}$Ce$_{x}$CuO$_{4}$ in
Figure \ref{EDCSummary}, summarizing our findings regarding this
LEG. For La$_{2-x}$Sr$_{x}$CuO$_{4}$, the topology of low-lying
excitations is qualitatively similar to
Ca$_{2-x}$Na$_{x}$CuO$_{2}$Cl$_{2}$, and the spectra are likewise
taken from the $d_{x^2 - y^2}$ nodal line. At a doping
concentration of \emph{x} = 0.01, dispersive low-energy states are
observed with a $\Delta_{\mathrm{LEG}}$ = 9 meV. However, by
\emph{x} = 0.03, this LEG has closed, to within our experimental
resolution, and remains as such for higher concentrations, as
studied in detail in Ref. \onlinecite{Yoshida02} (which focuses on
the metallic behavior for \emph{x} $\geq$ 0.03). In the case of
lightly electron-doped Nd$_{2-x}$Ce$_{x}$CuO$_{4}$, the first
electron addition states appear as small electron pockets near
($\pi$,0) \cite{Armitage02}, in contrast to the hole-doped
cuprates. In this case, we have observed the gap along these
electron pockets with $\Delta_{\mathrm{LEG}}$ $\sim$ 16 meV for
\emph{x} = 0.025, while for \emph{x} = 0.04,
$\Delta_{\mathrm{LEG}}$ $\sim$ 8 meV, and closes to nearly zero by
\emph{x} = 0.08; at higher concentrations (\emph{x} $> 0.10$), the
Fermi surface crosses over to a hole pocket centered at ($\pi,
\pi$).

The measurement of a LEG is the canonical scheme by which
excitation gaps have been typically determined by photoemission
spectroscopy. It is difficult to ascertain the precise value of
any excitation gap from the measurement of the gap without \emph{a
priori} knowledge of the single-particle spectral function,
$\mathcal{A}(k,\omega)$, making lineshape modelling potentially
suspect. However, the LEG criterion has typically been successful
in identifying the $d$-wave gap in the superconducting cuprates
\cite{Shen93}, charge density wave (CDW) gaps \cite{Kidd00}, and
even small superconducting gaps in photoemission studies of
conventional BCS materials such as V$_3$Si, Nb, and Pb
\cite{Reinert00,Chainani00}. Furthermore, our observation of
finite LEGs in a wide variety of lineshapes and compounds suggests
that this effect is not a misidentification due to a peculiar
lineshape profile. Nevertheless, we should note that it is not
inconceivable that in particular special instances, an ungapped
spectral function may possibly give rise to a finite LEG in the
ARPES lineshape (e.g. Luttinger liquids). All results were
confirmed by multiple measurements on different sample batches. We
have also utilized the method of symmetrization where $I_{\mathrm{
sym}}(\mathbf{k}, \omega) = I(\mathbf{k}, \omega) + I(\mathbf{k},
-\omega)$, which has been demonstrated to be an effective
procedure for determining the presence of Fermi crossings
\cite{Mesot01}. The results obtained from this method were
qualitatively consistent with values obtained by taking the
leading edge midpoints of the spectra.

Particular care was taken to avoid any electrostatic charging, a
possible consideration due to the low-temperature insulating
tendencies of these lightly doped samples. No change in
$\Delta_{\mathrm{LEG}}$ was observed when the photon flux was
varied by a factor of 3 or greater. Also, the macroscopic sample
surface quality for the lower doping concentrations was found to
be comparable to more heavily doped samples, as determined from
inspection by optical microscope and laser reflection. Finally,
all three studied families are chemically pristine when undoped
and must be alloyed towards higher doping levels. Therefore, the
opening of this gap towards lower concentrations cannot be
associated with a degradation in crystal quality, as would be the
case for the bismuth based cuprates. We note that ARPES studies of
irradiated Bi$_2$Sr$_2$CaCu$_2$O$_{8+\delta}$ have also shown a
demonstrable effect of induced disorder on the low energy spectral
lineshape \cite{Vobornik00}.

\begin{figure}[t]
\includegraphics{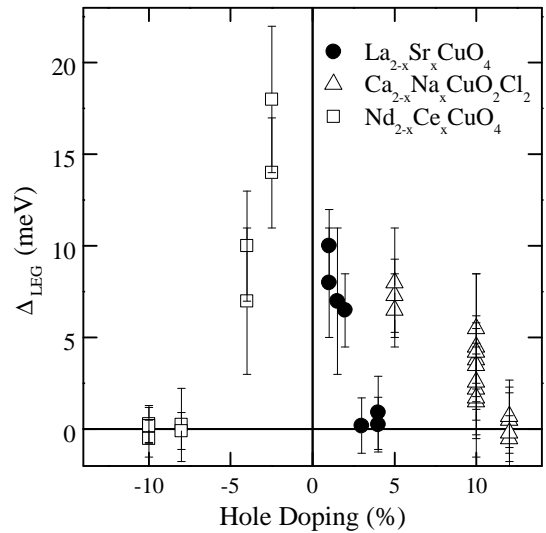} \vspace{0in}
\caption{Compilation of $\Delta_{\mathrm{LEG}}$ measurements taken
from all samples. Data from hole-doped samples were taken near
($\frac{\pi}{2}$, $\frac{\pi}{2}$) while data from electron-doped
samples were taken near ($\pi, 0$), as shown in the insets in
Figure \ref{EDCSummary}. All measurements were performed at 15 K.}
\label{GapSummary}
\end{figure}

The compositional and doping dependence of $\Delta_{\mathrm{LEG}}$
for all samples studied is summarized in Figure \ref{GapSummary}
and is shown to be reproducible over numerous measurements. In all
compounds, $\Delta_{\mathrm{LEG}}$ is largest at the lowest
concentrations and closes with increasing doping. Despite the
universal presence of this phenomenon, there exist obvious
differences in the detailed behavior of each particular compound.
In particular, the gap appears to close rapidly in
La$_{2-x}$Sr$_{x}$CuO$_{4}$.  Another intriguing point is that for
Ca$_{1.9}$Na$_{0.1}$CuO$_{2}$Cl$_{2}$, which has a
T$_{\mathrm{c}}$ onset of 13 K, there still exists a small but
finite $\Delta_{\mathrm{LEG}}$ $\sim$ 3 meV above 15 K, the base
temperature of our experiments. Interfamily variations in the
behavior of $\Delta_{\mathrm{LEG}}$ may not be unexpected, as many
other physical properties exhibit considerable material-specific
differences, including superconductivity and antiferromagnetism,
and may depend on factors such as the chemical composition and
crystal structure.

We now speculate on possible origins of the observed normal state
gap. One very important consideration is that disorder is
inherently manifest in the cuprates, as chemical substitution or
intercalation is necessary for introducing carriers. At
sufficiently low concentrations, the poor screening of these
impurities should cause a strong disorder potential which may
result in localization. The combination of disorder with
long-range Coulomb interactions can produce a depression in the
density of single-particle excitations at the chemical potential
known as a Coulomb or ``soft'' gap \cite{Efros75}, where the
presence of repulsive electron-electron interactions necessitates
a vanishing density of states at E$_{\mathrm{F}}$ to ensure
against an instability towards an excitonic ground state. The
existence of disorder and localization may also be consistent with
the reasonably broad peaks and edges, suggesting short lifetimes
and/or breaking of translational symmetry resulting in poorly
defined momentum eigenstates. Within this scenario, the reduction
of $\Delta_{\mathrm{LEG}}$ with doping can be explained by the
enhancement of screening. It should be emphasized that the
presence of a Coulomb gap even in the presence of disorder is
still a non-trivial result, since gapless insulating behavior can
also occur (i.e. weak localization); a Coulomb gap in the lightly
doped cuprates would be a clear indication of the strong
electron-electron interactions in these systems. It is also
possible that a Coulomb gap may exist in the lightly doped
cuprates without the aid of chemical disorder.

Recent STM studies have found considerable electronic
inhomogeneity in the cuprate superconductors
\cite{Pan01,KohsakaSTM02,Howald01}. In particular, results from
Kohsaka \emph{et al.} on Ca$_{1.92}$Na$_{0.08}$CuO$_{2}$Cl$_{2}$
have shown that this inhomogeneity persists to high energy scales
\cite{KohsakaSTM02}, implying that the distribution of carriers
varies strongly on nanometer length scales. A recent neutron
scattering study of lightly doped La$_{2-x}$Sr$_{x}$CuO$_{4}$
\cite{Matsuda02} also suggests the presence of electronic phase
separation below $x = 0.02$, close to where
$\Delta_{\mathrm{LEG}}$ vanishes. It is then possible that Coulomb
blockade effects in mesoscopic systems such as granular metals
\cite{Cuevas93} may be germane to this discussion. Photoemission
results from ultrathin granular Pb films \cite{Huang97} and
segmented one-dimensional systems \cite{Starowicz02}, have been
interpreted within such a framework. We note that the observed
values of $\Delta_{\mathrm{LEG}}$ appear rather small for Coulomb
blockade in the nanometer-sized patches proposed for the cuprates,
although additional effects such as the mutual screening of
patches and photohole relaxation \cite{Hovel98} may be mitigating
factors. Whether this inhomogeneity is driven solely by the
presence of chemical disorder or is an inherent property of the
pristine CuO$_2$ plane is still unclear. Nevertheless, both
chemical disorder and/or the presence of intrinsic electronic
inhomogeneity are plausible origins for a Coulomb gap which may
account for our results. We also note that the presence of
$\Delta_{\mathrm{LEG}}$ naturally reconciles the existence of an
insulating ground state, as determined from collective transport
properties, with the development of finite spectral weight near
E$_{\mathrm{F}}$ as measured from single-particle spectroscopy. It
is now clear that although the spectral intensity of excitations
near E$_{\mathrm{F}}$ grows as a function of doping, these
low-energy states are additionally gapped, resulting in a charge
and thermal insulator \cite{Iye92, Hawthorn03}.

Another intriguing possibility is that this may represent a
signature of an alternate phase of matter. It has been proposed
that the exotic normal state properties of the heavily underdoped
cuprates may signify the presence of a competing order, such as a
staggered flux phase \cite{Affleck88} or charge/spin stripes
\cite{Tranquada95}. It has also been established from neutron
scattering that spin density wave (SDW) order exists in
La$_{1.6-x}$Nd$_{0.4}$Sr$_{x}$CuO$_{4}$ \cite{Tranquada95} and
La$_{2-x}$Sr$_{x}$CuO$_{4+\delta}$ \cite{Wells97,Yamada98}. The
fact that we have consistently observed this effect in multiple
families suggests that if $\Delta_{\mathrm{LEG}}$ is due to a
competing order, this order should be generic to the cuprate
superconductors. In a competing order scenario, the doping
dependence in Figure \ref{GapSummary} suggests that the strength
of the competing phase decreases rapidly as a function of doping,
similar to the behavior of the pseudogap. We note that the
presence of chemical or electronic disorder does not necessarily
preclude the existence of a competing order. Future experiments
may help to clarify this situation. For instance, a systematic
study of $\Delta_{\mathrm{LEG}}$ with increasing chemical
impurities, such as Zn or Ni substitution, or induced disorder
\cite{Vobornik00}, would elucidate the effects of disorder on this
gap, and help to distinguish between a soft gap or a competing
order scenario.

In summary, we have presented ARPES results revealing the
existence of a finite gap over the entire Brillouin zone of the
lightly doped cuprates in the low-temperature normal state. This
phenomenon was observed in both electron and hole-doped cuprates
and was found to decrease as a function of carrier doping. We
believe this effect is one of the keys underlying the novel
superconductor-insulator transition in the lightly doped region of
the phase diagram and may represent electronic
inhomogeneity/disorder effects or a competing order parameter in
the lightly doped regime. It is hoped that these results will spur
future activity into developing a better understanding of the
properties of the lightly doped cuprates.

K.M.S. and Z.X.S. would like to thank F. Baumberger, R.B.
Laughlin, P.A. Lee, R.S. Markiewicz, N. Nagaosa, and S.C. Zhang
for insightful discussions. SSRL is operated by the DOE Office of
Basic Energy Science Divisions of Chemical Sciences and Material
Sciences. K.M.S. acknowledges SGF and NSERC for their support. The
ARPES measurements at Stanford were also supported by NSF
DMR0071897, ONR N00014-98-1-0195, and DOE Contract No.
DE-AC03-76SF00515. The Stanford crystal growth was supported by
DOE Contracts No. DE-FG03-99ER45773 and No. DE-AC03-76SF00515, by
NSF CAREER Award No. DMR-9985067, and by the A.P. Sloan
Foundation.


\end{document}